\documentclass[twocolumn,pra,showpacs,floatfix,citeautoscript,nofootinbib,superscriptaddress]{revtex4-1}
\usepackage{amsbsy}
\usepackage{latexsym,epsfig,graphicx}
\usepackage{dcolumn}
\usepackage{graphicx}
\usepackage{subfigure}
\usepackage{lipsum}
\usepackage{comment}
\usepackage{color}
\usepackage[colorlinks,urlcolor=blue,citecolor=blue]{hyperref}

\begin{document}

\title{Interacting spin-orbit-coupled spin-1 Bose-Einstein condensates}
\author{Kuei Sun}
\thanks{Authors with equal contributions.}
\affiliation{Department of Physics, The University of Texas at
Dallas, Richardson, Texas 75080-3021, USA}
\author{Chunlei Qu}
\thanks{Authors with equal contributions.}
\affiliation{Department of Physics, The University of Texas at
Dallas, Richardson, Texas 75080-3021, USA} \affiliation{INO-CNR
BEC Center and Dipartimento di Fisica, Universit\`a di Trento,
38123 Povo, Italy}
\author{Yong Xu}
\affiliation{Department of Physics, The University of Texas at
Dallas, Richardson, Texas 75080-3021, USA}
\author{Yongping Zhang}
\affiliation{Quantum Systems Unit, OIST Graduate University, Onna,
Okinawa 904-0495, Japan}
\author{Chuanwei Zhang}
\thanks{Corresponding author: chuanwei.zhang@utdallas.edu}
\affiliation{Department of Physics, The University of Texas at
Dallas, Richardson, Texas 75080-3021, USA}
\thanks{Corresponding author. chuanwei.zhang@utdallas.edu}

\begin{abstract}
The recent experimental realization of spin-orbit (SO) coupling
for spin-1 ultracold atoms opens an interesting avenue for
exploring SO-coupling-related physics in large-spin systems, which
is generally unattainable in electronic materials. In this paper,
we study the effects of interactions between atoms on the ground
states and collective excitations of SO-coupled spin-1
Bose-Einstein condensates (BECs) in the presence of a spin-tensor
potential. We find that ferromagnetic interaction between atoms
can induce a stripe phase exhibiting in-phase or out-of-phase
modulating patterns between spin-tensor and zero-spin-component
density waves. We characterize the phase transitions between
different phases using the spin-tensor density as well as the
collective dipole motion of the BEC. We show that there exists a
double maxon-roton structure in the Bogoliubov-excitation
spectrum, attributed to the three band minima of the SO-coupled
spin-1 BEC.
\end{abstract}

\maketitle

\vspace{-0.3cm}
\section{Introduction}\label{sec:introduction}
\vspace{-0.3cm}

Ultracold atoms have become a versatile platform for exploring
quantum matter in the presence of a variety of gauge
fields~\cite{Dalibard2011,Galitski2013,Goldman2014}. One important
breakthrough in this direction is the recent experimental
synthesis of coupling between a particle's spin and linear
momentum, or spin-orbit (SO) coupling, for pseudo-spin-half
ultracold
atoms~\cite{Lin2011,Zhang2012b,Qu2013a,Olson2014,Hamner2014,Wang2012,Cheuk2012,Williams2013,Huang2015}
via the light--atom interaction~\cite{Higbie2002,Spielman2009}.
Such synthetic SO coupling emulates the Rashba and Dresselhaus SO
coupling for electrons in solids that plays a crucial role for
many condensed-matter phenomena~\cite{Xiao2010,Qi2011}. Because of
the coupled spin and momentum, many interesting and exotic
SO-coupling-related phenomena have been theoretically proposed and
experimentally observed~\cite
{Stanescu2008,Wang2010,Wu2011,Ho2011,Zhang2012a,Hu2012,Ozawa2012,Li2012a,Gong2011,Hu2011,Yu2011,Qu2013b,Zhang2013b,Lin14}.

While electrons have spin-half, pseudospin of atoms, defined by
their hyperfine states, could have higher spins. For instance,
spin-1 Bose-Einstein condensates (BECs) have been widely studied
in ultracold atomic
gases~\cite{Ho1998,Ohmi1998,Stenger1998,Stamper-Kurn2013}. In this
context, the recent experimental realization of SO coupling for
spin-1 BECs through Raman coupling among three hyperfine
states~\cite{Campbell2015} (see Fig.~\ref{fig:f0}) or with the use
of a gradient magnetic field~\cite{Luo2015} provides an
interesting avenue for exploring SO-coupling-related physics in
high-spin systems. In the experiment of Ref.~\cite{Campbell2015},
the SO coupling is generated together with tunable
transverse-Zeeman and spin-tensor potentials. While the
transverse-Zeeman potential plays the same role as in the
spin-half system \cite{Higbie2002,Spielman2009}---to topologically
change the band structure by opening a gap, the spin-tensor
potential has fundamentally different rotation properties and acts
only in spin-1 (or higher-spin) space. Therefore, the system
enables exploration of spin-tensor-related physics in the
SO-coupled superfluid. The interplay between SO coupling and both
potentials leads to a rich single-particle band
structure~\cite{Lan2014}, which characterizes quantum phases with
itinerant magnetism and different types of phase transitions that
have been experimentally observed~\cite{Campbell2015}. However,
interaction effects on both ground state phase diagrams and
collective excitations in such spin-1 system have been largely
unexplored.

In this paper, we study the effects of density-density and
spin-spin interactions, two natural elements in spinor
BECs~\cite{Ho1998}, on the ground-state phase diagrams and
elementary excitations. Our results are based on the
variational-wavefunction approach and numerical simulation of
Gross-Pitaevskii equation (GPE), which agree with each other. The
main results are as follows:

\begin{figure}[b]\vspace{-0.3cm}
\centering
\includegraphics[width=8.6cm]{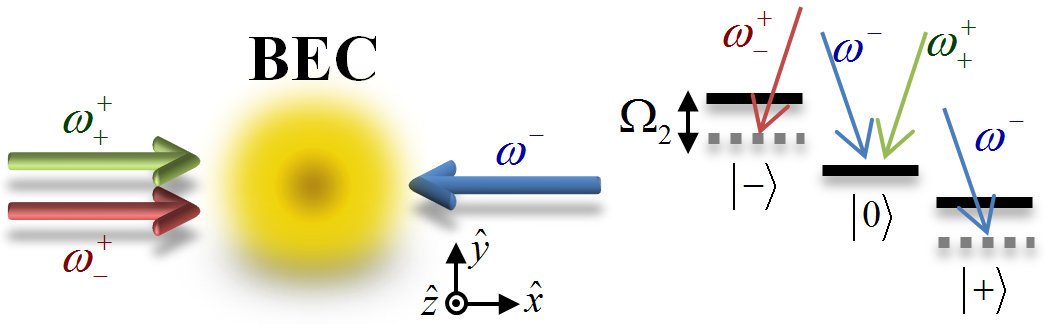} \vspace{-0.3cm}
\caption{(Color online) Left: The scheme to generate SO coupling
in a spin-1 BEC with three Raman lasers. Right: Raman transitions
between three hyperfine states with detuning $\Omega _{2}$, which
controls the spin-tensor potential.} \label{fig:f0}
\vspace{-0.3cm}
\end{figure}

 (1) The ground-state phase diagrams are obtained
for different parameters. Particularly, there exists an
interaction-induced stripe phase exhibiting in-phase or
out-of-phase modulations between spin-tensor and
zero-spin-component density waves. The size of the stripe-phase
region is proportional to the ferromagnetic interaction. The
stripe phase is the coherent superposition of components from
three band minima and possesses both crystalline and superfluid
properties, resembling the supersolid phases \cite{Pomeau1994}.

(2) The phase transitions (transition order and change of properties) are
characterized using the spin-tensor density, the time-of-flight momentum
distribution, and the period of collective dipole motions of the system.

(3) We reveal rich roton phenomena in the Bogoliubov-excitation
spectrum, including symmetric/asymmetric double-roton structures
and the roton gap closing at the stripe-phase boundary. Such
double-roton structures have not been found in widely studied
superfluids such as liquid $^{4}$He~\cite{Ozeri2005,Leggett2006},
BECs~with long range interactions \cite{Lahaye2009,Mottl2012}, and
SO-coupled spin-half BECs~\cite{Khamehchi2014,Ji2015,Ha2015},
which possess single-roton excitations.

\vspace{-0.3cm}
\section{Model and Hamiltonian}\label{sec:model}
\vspace{-0.3cm}

As illustrated in Fig.~\ref{fig:f0}, we consider the experimental
scheme of Ref.~\cite{Campbell2015}, which employs three Raman
lasers with wave vector $k_{\mathrm{R}}$ to generate SO coupling
in a spin-1 BEC. The pair of counterpropagating beams $\omega
_{-}$ and $\omega _{+}^{+}$ ($\omega _{-}^{+}$) induces a
two-photon Raman transition between two atomic hyperfine states
$|0\rangle $ and $|+\rangle $ ($|-\rangle $) and imparts
2$k_{\mathrm{R}}$ recoil momentum to the atoms. Following similar
discussion and modeling for the system at the single-particle
level in Refs.~\cite{Campbell2015,Lan2014,Campbell2015b}, we
consider the noninteracting Hamiltonian in pseudo-spin-1 basis
$\Psi ={\left( {\begin{array}{ccc} {\psi _{+}} & {\psi _{0}} &
{\psi _{-}}
\end{array} }\right) ^{T}}$ as
\begin{equation}
\tilde{H}_{0}=-\frac{{{\hbar ^{2}}{\nabla
^{2}}}}{{2m}}+{{\mathbf{\tilde{ \Omega}}}_{\mathbf{1}}}(x)\cdot
\mathbf{F}+{{\tilde{\Omega}}_{2}}F_{z}^{2}, \label{eq:Ham1}
\end{equation}
where an effective magnetic field
${{\mathbf{\tilde{\Omega}}}_{\mathbf{1}}}
(x)={{\tilde{\Omega}}_{1}}[\cos (2k_{\mathrm{R}}x)\hat{x}-\sin
(2k_{\mathrm{R }}x)\hat{y}]$, associated with the beam intensity,
couples to the hyperfine states represented by spin-1 Pauli
matrices $\mathbf{F}={\left( {
\begin{array}{ccc}
{F_{x},} & {F_{y},} & {F_{z}}
\end{array}
}\right) }$, $\tilde{\Omega}_{2}$ is an effective detuning for
$|\pm \rangle $ (hence taking a spin-tensor form $F_{z}^{2}$), and
$m$ is the atomic mass. Since
${{\mathbf{\tilde{\Omega}}}_{\mathbf{1}}} (x)$ has effects only in
the spatial $x$ direction, one can assume that the ground-state
wavefunction in the $y$ and $z$ directions remains the case
without it. After unitary transformation $\psi _{\pm }\rightarrow
\psi _{\pm }e^{\pm 2ik_{\mathrm{R}}x}$ and integration over $y$
and $z$ degrees of freedom, we obtain a Hamiltonian in momentum
and energy units $\hbar k_{\mathrm{R}}$ and $\hbar
^{2}k_{\mathrm{R}}^{2}/2m$, respectively, as
\begin{equation}
H_{0}=-\partial _{x}^{2}-4i{\partial _{x}} \otimes
{F_{z}}+({\Omega _{2}}+4)F_{z}^{2}+ {\Omega _{1}}{F_{x}},
\label{eq:Ham2}
\end{equation}
where $\Omega _{1,2}$ are dimensionless variables controlling the
transverse-Zeeman and spin-tensor potentials, respectively. The
term $-i\partial _{x} \otimes F_{z}=p_{x} \otimes F_{z}$
represents the spin--linear-momentum coupling.

The ground-state properties of $H_{0}$ can be characterized by
considering the minima in the lowest energy band,
\begin{equation}
{E_{k}}={k^{2}}-\sqrt[^{3}]{{A}_{k}^{\prime
}/54}-{A_{k}}\sqrt[^{3}]{2/(27{{ \ A_{k}^{\prime }}})}+2{A_{0}/3},
\label{eq:dispersion}
\end{equation}
where $A_{0}=\Omega _{2}+4$, ${A_{k}}=48{k^{2}}+A_{0}^{2}+3\Omega
_{1}^{2}$, and ${A_{k}^{\prime }}={A_{0}}{A_{k}^{\prime \prime
}}+\sqrt{-4A_{k}^{3}+{ A_{0}^{2}}{{A_{k}^{\prime \prime }}^{2}}}$
with ${A_{k}^{\prime \prime }} =-288{k^{2}}+2A_{0}^{2}+9\Omega
_{1}^{2}$. The structure of $E_{k}$ can exhibit (A) one local
minimum at $k=0$ [right inset in Fig.~\ref{fig:f1} (a)], (B) two
at $k=\pm k_{0}\neq 0$ (bottom inset), or (C) three at $k=0,\pm
k_{0}$ (top two insets) as $\Omega _{1,2}$ vary. The ground state
always stays at $k=0$ in the region $\Omega _{2}>0$ and can
undergo a phase transition between $0$ and $\pm k_{0}$ in $\Omega
_{2}\leq 0$. Along the phase boundary in the $\Omega _{1}$-$\Omega
_{2}$ plane, there is a triple point $(\Omega _{1}^{\ast },\Omega
_{2}^{\ast })$ that separates two types of transitions: a
first-order transition in $\Omega _{1}<\Omega _{1}^{\ast }$ upon
which structure (C) remains and $k$ suddenly jumps (top two
insets) and a second-order transition upon which the structure
evolves between (A) and (B) and $k$ continuously changes (bottom
two insets). The boundaries of first-order and second-order
transitions meet the conditions $E_{0}=E_{\pm k_{0}}$ and
$\partial _{k}^{2}E_{0}=0$, respectively, which represent a
monotonically decreasing curve in the $\Omega _{1}$-$\Omega _{2}$
plane. The triple point takes place at the merging of the three
local minima $\pm k_{0}\rightarrow 0$ of structure (C), which
gives $(\Omega _{1}^{\ast },\Omega _{2}^{\ast })=(4.805,-1.666)$
and a flat low-energy behavior $E_{k}\sim
0.0165k^{6}$~\cite{Triple_point}.

\vspace{-0.3cm}
\section{Variational calculation}\label{sec:variational}
\vspace{-0.3cm}

For a realistic BEC, the energy density of the system can be
expressed as
\begin{equation}
\varepsilon =\frac{1}{V}\int {dx{\Psi ^{\dag }}\left[
{{H_{0}}+\frac{{c_{0}} }{2}{\Psi ^{\dag }}\Psi
+\frac{{c_{2}}}{2}\left( {{\Psi ^{\dag }}\mathbf{F} \Psi }\right)
\cdot \mathbf{F}}\right] \Psi },  \label{eq:energy}
\end{equation}
where $V$ is the system volume and $c_{0,2}$ describe
density-density and spin-spin interaction strengths. The wave
function is normalized as $V^{-1}\int {dx{\Psi ^{\dag }}\Psi }=1$
such that $c_{0,2}$ are proportional to the particle density
$N/V$. Note that $c_{0,2}$ could also be tuned through Feshbach
resonances~\cite{Chin10}. To obtain the ground state, we adopt a
variational ansatz,
\begin{equation}
\Psi =\left\vert {C_{0}}\right\vert {\chi
_{0}}+\sum\nolimits_{s=\pm }{\left\vert {C_{s}}\right\vert {\chi
_{s}}{e^{is(kx+{\alpha _{s}})}}}, \label{eq:ansatz}
\end{equation}
where the spin components are of the form ${\chi _{-}}(\theta
,\phi )={\left( {\begin{array}{ccc} {\cos \theta \cos \phi } & {\
-\sin \theta } & {\cos \theta \sin \phi }
\end{array}
}\right) ^{T}}$, ${\chi _{0}}={\chi _{-}}(\frac{\pi }{2}-\theta'
,\frac{\pi }{ 4})$, and ${\chi _{+}}={\chi _{-}}(\theta ,\frac{\pi
}{2}-\phi )$. The normalization condition gives
$\sum\nolimits_{s=0,\pm }{{{\left\vert {C_{s}} \right\vert
}^{2}}}=1$. Inserting Eq.~(\ref{eq:ansatz}) into
Eq.~(\ref{eq:energy}), we obtain $\varepsilon $ as a functional of
seven variational parameters $k$, $|C_{\pm }|$, $\theta $,
$\theta' $, $\phi $, and $\alpha =\alpha _{-}-\alpha _{+}$ (see
Appendix \ref{sec:functional}), which are generally different from
their single-particle values \cite{Lan2014,Natu2015}. The ground
state is computed through the minimization of $\varepsilon $. We
also calculate the ground state by solving the GPE using imaginary
time evolution and find good agreement between both methods.

In terms of the variational variables, we obtain the spin
polarization $\left\langle {F_{z}}\right\rangle =({\left\vert
{C_{+}}\right\vert ^{2}}-{\left\vert {C_{-}}\right\vert
^{2}}){\cos ^{2}}\theta \cos 2\phi $ and spin tensor $\left\langle
{F_{z}^{2}}\right\rangle =({\left\vert {C_{+}} \right\vert
^{2}}+{\left\vert {C_{-}}\right\vert ^{2}})\cos^2\theta +
{\left\vert {C_{0}}\right\vert ^{2}}{\sin ^{2}}\theta' $, which
are directly related to the spin-resolved density profiles
${V^{-1}}\int {dx({{\rho_+}\mp {\rho_-}})} $ with $\rho_s =
|\psi_s|^2$, respectively, and can hence be measured in the
time-of-flight experiment. The minimization of energy with respect
to $k$, i.e., $\partial \varepsilon /\partial k=0$, leads to the
ground-state spin polarization associated with momentum $\langle
{F_{z}}\rangle =-\frac{k}{2}({\left\vert {C_{+}}\right\vert
^{2}}-{\left\vert {C_{-}}\right\vert ^{2}})$, which is a result of
SO coupling.

\vspace{-0.3cm}
\section{Interacting phase diagram}\label{sec:phase_diagram}
\vspace{-0.3cm}

The variational ansatz characterizes three quantum phases: (I) the
uniform phase with a constant wavefunction at $k=0$, and hence
$\langle {F_{z}}\rangle =0$; (II) the plane-wave phase with $k\neq
0$, one of $|C_{\pm }|$ equal to $1$, and hence $\langle
{F_{z}}\rangle \neq 0$; and (III) the stripe phase with $k\neq 0$
and at least two of $|C_{0,\pm }|$ are not $0$. The plane-wave
phase has two degenerate states with opposite momentum and hence
opposite spin polarization. Below we present only the positively
polarized plane-wave state for convenience. The stripe phase
exhibits spatial density modulation $\rho
(x)=\sum\nolimits_{s=0,\pm } \rho_s$ with periods determined by
the superposition of the uniform and plane-wave states. The
transition between phases is first order (second order) if
$\frac{\partial \varepsilon }{\partial \Omega _{2}}=\langle
F_{z}^{2}\rangle $ ($\frac{\partial ^{2}\varepsilon }{\partial
\Omega _{2}^{2}}=\frac{\partial \langle F_{z}^{2}\rangle
}{\partial \Omega _{2}}$)~\cite{Hellmann-Feynman} displays
discontinuity as $\Omega _{2}$ varies across the phase boundary.
Another polarization $\langle F_{x}\rangle =\frac{\partial
\varepsilon }{
\partial \Omega _{1}}$ can also indicate the type of phase
transitions but is less experimentally accessible. Note that the
behavior of $\langle F_{z}\rangle $ is not directly related to the
energy derivatives in the $\Omega_1$-$\Omega_2$ plane. In fact it
can not tell the transition involving the stripe phase, as we will
show later.

\begin{figure}[t]
\centering
\includegraphics[width=8.6cm]{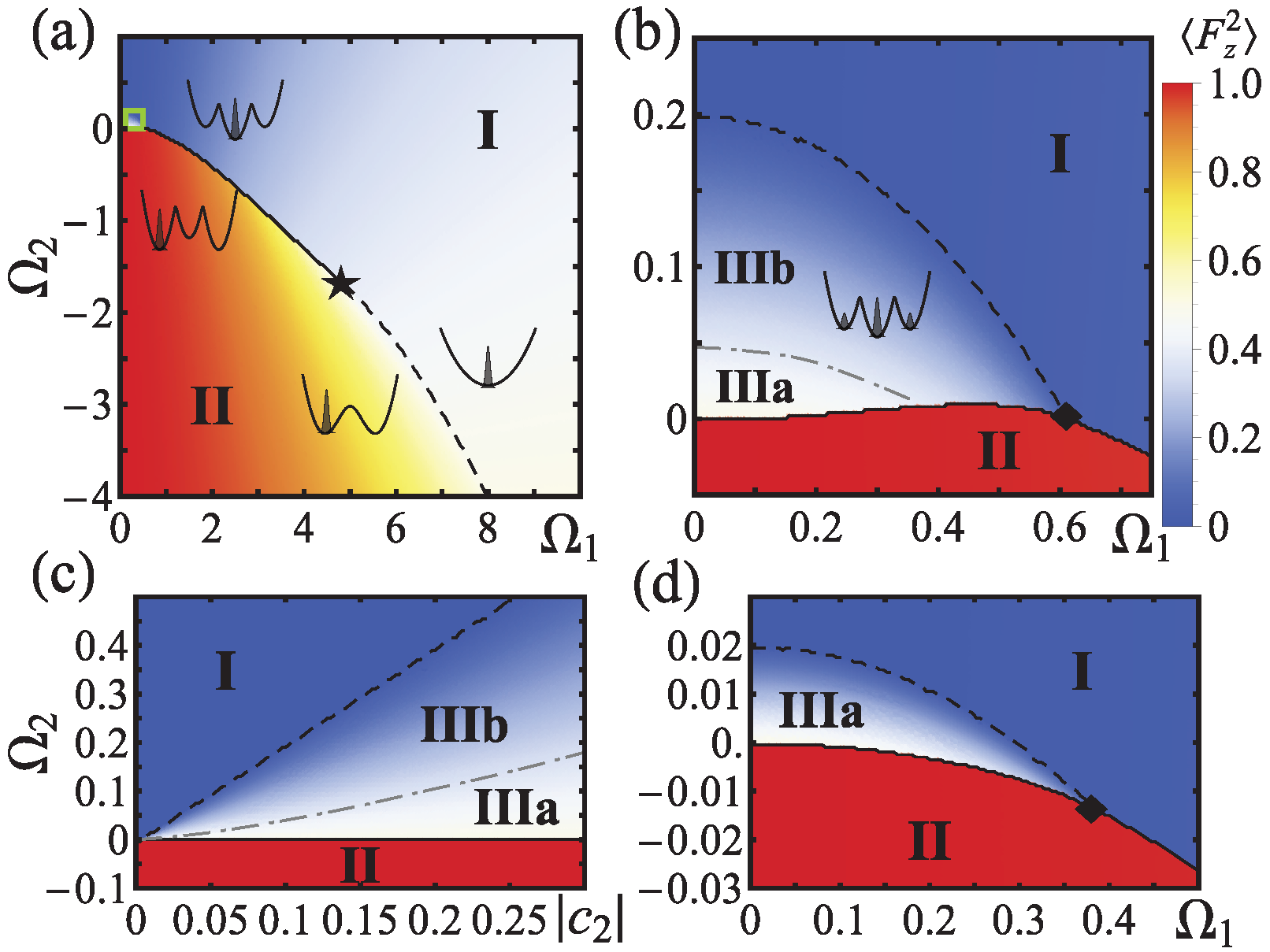} \vspace{-0.3cm}
\caption{(Color online) (a) Ground-state phase diagram in the
$\Omega _{1}$-$\Omega _{2}$ plane for $(c_{0},c_{2})=(5,-0.1)$.
(b) Closeup of the framed region in panel (a). Regions I, II, and
IIIa/IIIb, represent uniform, plane-wave, and stripe phases,
respectively. The solid, dashed, and dot-dashed curves show
first-order-transition, second-order-transition, and crossover
boundaries, respectively. The star (diamond) sign denotes the
triple (tricritical) point. The insets show schematic
single-particle band structures and the BEC's occupation in the
momentum space. (c) Phase diagram in the $|c_{2}|$-$\Omega _{2}$
plane for $c_{0}=5$ and $\Omega _{1}=0.1$. (d) Phase diagram in
the $\Omega _{1}$-$\Omega _{2}$ plane for
$(c_{0},c_{2})=(0.5,-0.01)$.} \label{fig:f1} \vspace{-0.3cm}
\end{figure}

Figure \ref{fig:f1}(a) shows the ground-state phase diagram in the
$\Omega _{1}$-$\Omega _{2}$ plane at $(c_{0},c_{2})=(5,-0.1)$.
Outside the small framed region, we find that such small spin-spin
interaction $c_{2}$ has little effects on the phase diagram. The
quantum phases and phase transitions can still be well described
by the single-particle energy band (see insets), except that the
system always stays in one of the minima given the degeneracy due
to the density-density interaction $c_{0}$, in contrast to the
noninteracting ground state, which can be arbitrary superposition
of degenerate minima. As a result, the boundaries of first-order
(solid curve) and second-order (dashed) transitions between the
uniform (I) and plane-wave (II) phases as well as the place of
triple point (star sign) show indiscernible difference from the
noninteracting case. When $c_{2}$ becomes stronger to $-2$, it
enlarges region II but does not qualitatively change the I--II
boundary.

Figure \ref{fig:f1}(b) zooms in the framed region in
Fig.~\ref{fig:f1}(a). We see the appearance of a stripe phase III
sandwiched by I and II. The III--II and III--I transition
boundaries are first order (solid curve) and second order
(dashed), respectively, and meet at a tricritical point
$(0.6,0.005)$ (diamond sign). The stripe phase represents
co-occupancy of the three states of the form
$|C_{0}|>|C_{+}|=|C_{-}|>0$ (see insets), so its direct evidence
would be a symmetric three-peak structure at $0$ and $\pm k$ in
the time-of-flight experiment. Such a superposition leads to
spatially modulated $\rho$ and spin-tensor density $\rho_+ +
\rho_-$ with period $2\pi /k$, while the spin polarization $\rho_+
- \rho_- =0$ remains uniform. The stripe phase shows two patterns:
$\rho_+ + \rho_-$ and $\rho_0$ have (IIIa) in-phase modulations
(coincident maxima and minima) or (IIIb) out-of-phase modulations.
The IIIb pattern results from a stronger interband mixing due to
the interaction. The amplitude of $\rho_0$ smoothly suppresses to
zero on the IIIa--IIIb crossover boundary (dot-dashed curve) and
changes sign across the boundary.

We turn to study the effect of interaction strength on the phase
boundaries. Figure \ref{fig:f1}(c) shows the phase diagram in the
$|c_{2}|$-$\Omega _{2}$ plane ($c_{2} < 0$ remains ferromagnetic)
for $c_{0}=5$ and $\Omega _{1}=0.1$. We see that the stripe-phase
region linearly increases with $|c_{2}|$. For weaker interaction
$(c_{0},c_{2})=(0.5,-0.01)$, we find that IIIb disappears and the
II--III boundary monotonically decreases from $\Omega_2=0$ as in
the single-particle case [see Fig.~\ref{fig:f1}(d)], but the
linear increase of the stripe region with $|c_{2}|$ remains. Since
$|c_{2}|$ is proportional to the average particle density, we
expect the stripe phase to be more attainable in a dense system. A
$^{87}$Rb system with density $10^{15} \mathrm{{ cm}^{-3}}$,
$s$-wave scattering length $100.48a_{0}$ ($a_{0}$ is the Bohr
radius), and Raman-laser wavelength $800$ nm corresponds to
$c_{0}=2.2$ and $c_{2}=-0.01$, which predict the stripe phase
within $\tilde{\Omega}_{1}<0.23E_{R}=814.2h\times \mathrm{Hz}$ and
$\tilde{\Omega}_{2}<0.02E_{R}=70.8h\times \mathrm{Hz}$. For
$^{133}$Cs, the difference between intraspin and interspin
interactions is easily tunable due to a broad Feshbach resonance
of the intraspin scattering~\cite{Chin10}, so $|c_{2}|$ can be
enhanced without increasing the density.

\begin{figure}[t]
\centering
\includegraphics[width=8.6cm]{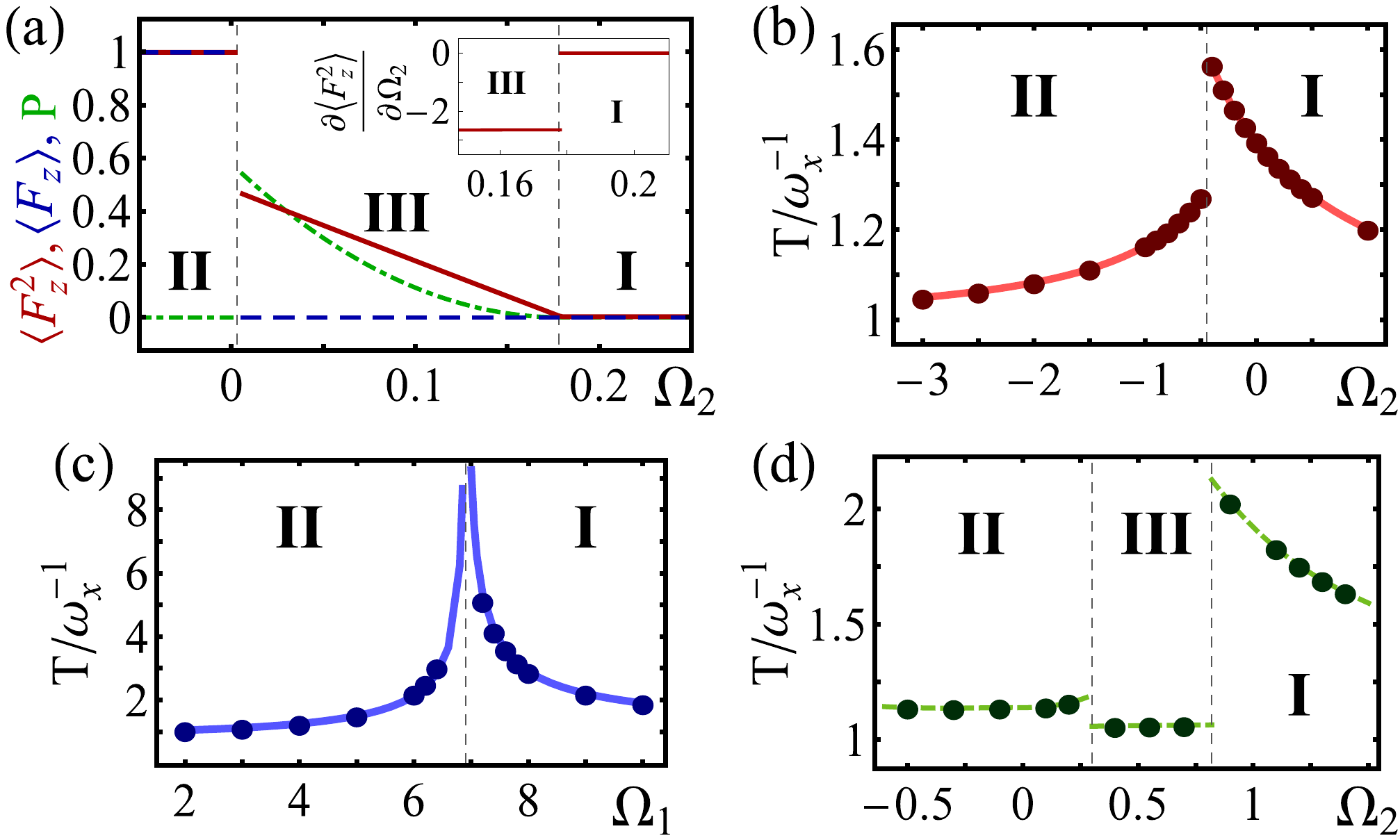} \vspace{-0.3cm}
\caption{(Color online) (a) $\langle F_z^2 \rangle$ (solid curve),
$\langle F_z \rangle$ (dashed), and $P=10|C_+^2 C_-^2|$
(dot-dashed) vs $\Omega_2$ along path $\Omega_1=0.2$ in
Fig.~\ref{fig:f1}(b). The inset shows $\frac{\partial \langle
F_z^2 \rangle}{\partial \Omega_2}$. (b) [(c)] Dipole oscillation
periods $T$ along path $\Omega_1=2$ ($\Omega_2=-3$) across a
first-order (second-order) transition. Filled circles (solid
curves) are obtained from the GPE (effective-mass method). (d) $T$
across the stripe phase along path $\Omega_1=1.3$ (with
thick-dashed fitting curves). Phases are labeled in the
corresponding regions separated by dashed lines.} \label{fig:f2}
\vspace{-0.3cm}
\end{figure}

To show the system behavior upon the phase transitions, we plot
several observables in Fig.~\ref{fig:f2}(a), including $\langle
F_{z}^{2}\rangle $ (solid curve), $\langle F_{z}\rangle $
(dashed), and combined occupancy $P=|C_{+}^{2}C_{-}^{2}|\times 10$
(dot-dashed) vs $\Omega _{2}$ along path $\Omega _{1}=0.2$ in
Fig.~\ref{fig:f1}(b). The discontinuity in $\langle
F_{z}^{2}\rangle $ [$\frac{\partial \langle F_{z}^{2}\rangle
}{\partial \Omega _{2}}$ (see inset)] indicates the II--III
(III--I) transition to be first (second) order. The $\langle
F_{z}\rangle $ curve can also indicate the II--III transition but
not III--I. The stripe phase has $\langle F_{z}\rangle =0$, the
same as the uniform phase, but exhibits nonzero co-occupancy
$P>0$. This underlines the insufficiency to characterize the
interacting systems with only spin polarization.

\vspace{-0.3cm}
\section{Dipole oscillation} \label{sec:dipole}
\vspace{-0.3cm}

We study dipole collective modes of a trapped BEC as another
experimental probe~\cite{Zhang2012b} for the phase transitions. We
consider an experimental setup of $5\times 10^{4}$ $^{87}$Rb atoms
with the aforementioned scattering length and Raman-laser
wavelength and anisotropic trapping frequencies $\omega
_{x}=\omega _{y}=0.23\omega _{z}=2 \pi \times 33$ Hz, producing an
effective two-dimensional system in our simulation (where we
integrate out only the $z$ degrees of freedom). After an initial
displacement of $1.26$ $\mu $m in the $x$ direction, we compute
the BEC's periodic motion by numerically solving the
time-dependent GPE and record the oscillation period $T$. In our
simulation, we do not see effects of the $y$ degrees of freedom on
$T$. We first study the transition between phases I and II. Figure
\ref{fig:f2}(b) [(c)] shows that $T$ exhibits a discontinuity
(diverges) upon the first-order (second-order) transition along
path $\Omega _{1}=2$ ($\Omega _{2}=-3$) in the $\Omega
_{1}$-$\Omega _{2}$ plane. Such behaviors well match the
effective-mass approximation~\cite{Chen2012} (see solid curves),
in which the dipole motion is considered as a semi-classical
simple harmonic oscillator subject to the energy dispersion
$E_{k}$ of Eq.~(\ref{eq:dispersion}). The displacement $x(t)$ and
momentum $k(t)$ obey the equation of motion $\partial
_{t}k=-\omega _{x}^{2}x$ and $\partial _{t}x=\partial
E_{k}/\partial k$, resulting in the period $T\propto
\sqrt{m_{\mathrm{eff}}}$ with effective mass
$m_{\mathrm{eff}}=(\partial ^{2}E_{k}/\partial k^{2})^{-1}$
defined by the band curvature. Therefore, the divergence of $T$
upon the second-order transition comes from the band flatness
$\partial ^{2}E_{k}/\partial k^{2}=0$. By expanding the curvature
around a second-order transition point $(\Omega _{1}^{c},\Omega
_{2}^{c})$, we obtain the critical behavior $T\propto |\Omega
_{1(2)}-\Omega _{1(2)}^{c}|^{-1/2}$ at fixed $\Omega _{2(1)}$,
with the critical exponent consistent with the spin-half
case~\cite{Li2012b}. For the stripe phase, we find that its dipole
oscillation period is lower than those of the other phases. To
reveal the salient feature of such a trend, we study a strongly
interacting system with $(c_{0},c_{2})=(10,-2)$ in a
one-dimensional trap along path $\Omega _{1}=1.3$. Figure
\ref{fig:f2}(d) shows the results from the GPE calculations (the
effective-mass approximation no longer fits here). We see a clear
drop in the period of the stripe phase compared with the nonstripe
phases, with the discontinuities matching the boundaries (dashed
lines).

\begin{figure}[b]\vspace{-0.3cm}
\centering
\includegraphics[width=6cm]{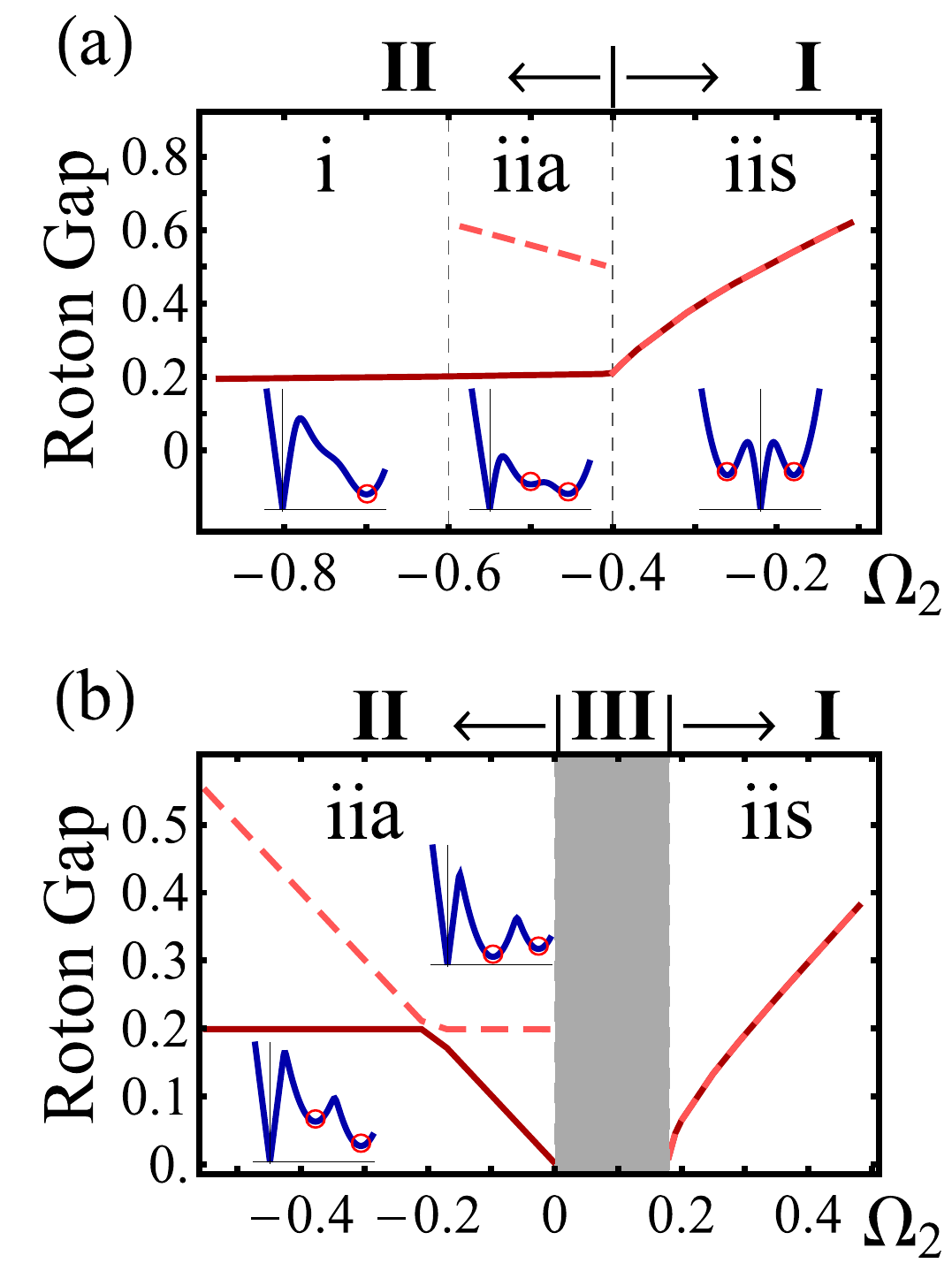} \vspace{-0.3cm}
\caption{(Color online) (a), (b) The lowest (solid curve) and
second-lowest (dashed curve) roton gaps vs $\Omega _{2}$ along
paths $\Omega _{1}=2$ in Fig.~\ref{fig:f1}(a) and $0.2$ in
Fig.~\ref{fig:f1}(b), respectively. Regions of different roton
structures (see text) are labeled and separated by dashed lines.
The corresponding ground-state phases are labeled above the
panels. Curves in the insets show schematic Bogoliubov-excitation
spectra ($x$ and $y$ axes represent momentum and energy,
respectively) in the corresponding regions. The roton modes are
circled. There are no roton excitations in the shaded region of
the stripe phase~\cite{Li2013}.} \label{fig:f3} \vspace{-0.3cm}
\end{figure}

\vspace{-0.3cm}
\section{Double rotons} \label{sec:rotons}
\vspace{-0.3cm}

The triple-well band structure of the spin-1 BEC leads to exotic
quasiparticle excitations that do not exist in previous spin-half
or long-range-interacting systems. We calculate the Bogoliubov
excitations for phases I and II by solving the Bogoliubov
equation, derived by linearizing the GPE with plane-wave-type
particle and hole perturbations (see Appendix
\ref{sec:Bogoliubov}). We first see that the excitation spectrum
is always gapless and linear in the small-momentum region,
presenting typical phonon modes in BECs. In a larger-momentum
region, we find a rich structure of rotons, i.e., quasistable
modes with zero group velocity. In Fig.~\ref{fig:f3}(a), we plot
the energy gaps of the lowest and second-lowest (if present)
rotons along path $\Omega _{1}=2$ in Fig.~\ref{fig:f1}(a). There
are two types of roton structures in phase II: (i) single roton
and (iia) asymmetric double rotons. In phase I, there are
symmetric and degenerate double rotons (iis) (see insets for
schematic spectra with marked rotons). Therefore, the sudden
change in the roton structure can be a signature for the II--I
transition. Figure \ref{fig:f3}(b) shows the results along path
$\Omega _{1}=0.2$. We see that when the system approaches to III
from II, the two roton gaps first cross each other. Then the one
at lower momentum keeps decreasing to zero at the II--III
boundary, while the other remains finite. If approaching from I,
the degenerate gaps both drop to zero at the I--III boundary. Such
gap closing is a signature of the transition to the stripe phase,
resembling to the transition to the supersolid phase that
possesses both crystalline and superfluid orders. The trend of the
reducing roton gap close to the stripe phase was experimentally
observed previously for the spin-half
system~\cite{Khamehchi2014,Ji2015}. The Bogoliubov excitations in
the stripe phase itself are more complicated and there are no
rotons~\cite{Li2013} (shaded region). Note that such double rotons
have not been found in the previously studied
superfluids~\cite{Ozeri2005,Leggett2006,Lahaye2009,Mottl2012,Khamehchi2014,Ji2015,Ha2015}.

\vspace{-0.3cm}
\section{Summary} \label{sec:summary}
\vspace{-0.3cm}

In summary, we have characterized the ground-state phase diagrams
and collective excitations in interacting SO-coupled spin-1 BECs
with spin-tensor potentials. Our results provide timely
predictions for ongoing experiments exploring SO-coupling-related
physics in higher-spin systems. One interesting extension would be
the investigation of spin-1 BECs with recently proposed
spin--orbital-angular-momentum
coupling~\cite{Sun2015,Demarco2015,Qu2015}, in which the strongly
interacting effects might significantly enlarge the stripe-phase
region.

\textbf{Acknowledgements}: We are grateful to P. Engels, L. Jiang,
and Z.-Q. Yu for helpful discussions. This work is supported by
ARO (W911NF-12-1-0334), AFOSR (FA9550-13-1-0045), and NSF
(PHY-1505496).

\onecolumngrid

\appendix
\vspace{-0.3cm}
\section{Variational energy functional} \label{sec:functional} \vspace{-0.3cm}
In this section we show details of the variational energy
functional density $\epsilon$ discussed in
Sec.~\ref{sec:variational}. We first rewrite the variational
wavefunction in Eq.~(\ref{eq:ansatz}) as
\begin{eqnarray}
\left( {\begin{array}{*{20}{c}}
{{\psi _ + }}\\
{{\psi _0}}\\
{{\psi _ - }}
\end{array}} \right) = \left| {{C_ - }} \right|\left( {\begin{array}{*{20}{c}}
{\cos \theta \cos \phi }\\
{ - \sin \theta }\\
{\cos \theta \sin \phi }
\end{array}} \right){e^{ - i(kx + {\alpha _ - })}} +
 \left| {{C_0}} \right|\left( {\begin{array}{*{20}{c}}
{\sin \theta '/\sqrt 2 }\\
{ - \cos \theta '}\\
{\sin \theta '/\sqrt 2 }
\end{array}} \right) + \left| {{C_ +}} \right|\left( {\begin{array}{*{20}{c}}
{\cos \theta \sin \phi }\\
{ - \sin \theta }\\
{\cos \theta \cos \phi }
\end{array}} \right){e^{i(kx + {\alpha _ + })}}.
\label{eq:variational_detail}
\end{eqnarray}
We then evaluate each term of the single-particle energy density
using Eq.~(\ref{eq:variational_detail}) as
\begin{eqnarray}
\left\langle { - \partial _x^2} \right\rangle  &=& \frac{1}{V}\int
{dx} \sum\limits_j {{{\psi }^*_j}( - \partial _x^2){\psi _j}} =
{k^2}\left( {{{\left| {{C_+}} \right|}^2} + {{\left| {{C_-}}
\right|}^2}} \right), \\
\left\langle {-4i{\partial _x}{F_z}} \right\rangle  &=&
\frac{1}{V}\int {dx} (-4i)({{ \psi }^*_ + }{\partial _x}{\psi _ +
} - {{\psi }^*_ - }{\partial _x}{\psi _ - })  = -4k{\cos ^2}\theta
\cos 2\phi \left( {{{\left| {{C_+}} \right|}^2} + {{\left| {{C_-}}
\right|}^2}} \right),\\
 \left\langle {F_z^2} \right\rangle  &=& \frac{1}{V}\int {dx\left(
{{{\left| {{\psi _ + }} \right|}^2} + {{\left| {{\psi _ - }}
\right|}^2}} \right)} = \left( {{{\left| {{C_ + }} \right|}^2} +
{{\left| {{C_ - }} \right|}^2}} \right){\cos ^2}\theta  + {\left|
{{C_0}} \right|^2}{\sin ^2}\theta ', \\
 \left\langle {{F_x}} \right\rangle  &=& \frac{1}{V}\int
{dx\frac{1}{{\sqrt 2 }}\left( {\psi _ + ^\dag {\psi _0} + \psi
_0^\dag {\psi _ - } + {\rm{H}}{\rm{.c}}{\rm{.}}} \right)} = -
\left( {{{\left| {{C_ + }} \right|}^2} + {{\left| {{C_ - }}
\right|}^2}} \right)\sin 2\theta \sin \left( {\phi  + \frac{\pi
}{4}} \right) - {\left| {{C_0}} \right|^2}\sin 2\theta ',
\end{eqnarray}
and obtain
\begin{eqnarray}
{\varepsilon _0} = \left\langle { - \partial _x^2 - 4i{\partial
_x}{F_z}} \right\rangle  + ({\Omega _2} + 4)\left\langle {F_z^2}
\right\rangle  + {\Omega _1}\left\langle {{F_x}} \right\rangle.
\label{eq:energy_density} \label{eq:E0}
\end{eqnarray}
For the interaction energy, we have
\begin{eqnarray}
{\varepsilon _{{c_0}}} &=& \frac{{{c_0}}}{2}\frac{1}{V}\int
{dx{{\left( {{{\left| {{\psi _ + }} \right|}^2} + {{\left| {{\psi
_0}} \right|}^2} + {{\left| {{\psi _ - }} \right|}^2}}
\right)}^2}} = \frac{{{c_0}}}{2} \Bigg\{ 1 + 2{\left| {{C_0}}
\right|^2}{\left[ {\cos \theta \sin \theta '\sin \left( {\phi  +
\frac{\pi }{4}} \right) + \sin \theta \cos \theta '} \right]^2}
\times \nonumber\\ && \left( {{{\left| {{C_ + }} \right|}^2} +
2\left| {{C_ + }} \right|\left| {{C_ - }} \right|\cos \alpha  +
{{\left| {{C_ - }} \right|}^2}} \right) +  2{\left| {{C_ + }}
\right|^2}{\left| {{C_ - }} \right|^2}{\left( {{{\cos }^2}\theta
\sin 2\phi  + {{\sin }^2}\theta } \right)^2} \Bigg\},
\label{eq:Ec0}
\end{eqnarray}
\begin{eqnarray}
{\varepsilon _{{c_2}}} = \frac{{{c_2}}}{2}\frac{1}{V}\int
{dx\sum\limits_{s = x,y,z} {{{\left( {\sum\limits_{ij =  + ,0, - }
{{{\psi }^*_i}{F_{s,ij}}{\psi _j}} } \right)}^2}} }  =
\frac{{{c_2}}}{2}\frac{1}{V}\int {dx\left[ {2{{\left| {{{\psi }^*_
+ }{\psi _0} + {{\psi }^*_0}{\psi _ - }} \right|}^2} + {{\left(
{{{\left| {{\psi _ + }} \right|}^2} - {{\left| {{\psi _ - }}
\right|}^2}} \right)}^2}} \right]}, \label{eq:Ec2}
\end{eqnarray}
with
\begin{eqnarray}
&&\frac{1}{V}\int {dx} {\left| {{{\psi }^*_ + }{\psi _0} + {{ \psi
}^*_0}{\psi _ - }} \right|^2} = \frac{1}{2}{\left[ {\left(
{{{\left| {{C_+}} \right|}^2} + {{\left| {{C_-}} \right|}^2}}
\right)\sin 2\theta \sin \left( {\phi  + \frac{\pi }{4}} \right) +
{{\left| {{C_0}} \right|}^2}\sin 2\theta '} \right]^2} + {\left|
{{C_+}} \right|^2}{\left| {{C_-}} \right|^2}{\sin ^2}2\theta +
\nonumber\\
&&  {\left| {{C_0}} \right|^2}\left( {{{\left| {{C_+}} \right|}^2}
+ {{\left| {{C_-}} \right|}^2} + 2\left| {{C_+}} \right|\left|
{{C_-}} \right|\cos {\alpha}} \right)\left[ {{{\cos }^2}\theta
{{\cos }^2}\theta ' + \frac{1}{2}\sin 2\theta \sin 2\theta '\sin
\left( {\phi  + \frac{\pi }{4}} \right) + {{\sin }^2}\theta {{\sin
}^2}\theta '} \right] , \label{eq:Ec2a}
\end{eqnarray}
\begin{eqnarray}
\frac{1}{V}\int {dx{{\left( {{{\left| {{\psi _ + }} \right|}^2} -
{{\left| {{\psi _ - }} \right|}^2}} \right)}^2}}  &=& {\left(
{{{\left| {{C_+}} \right|}^2} - {{\left| {{C_-}} \right|}^2}}
\right)^2}{\cos ^4}\theta {\cos ^2}2\phi + \nonumber\\
&& 2{\left| {{C_0}} \right|^2}{\cos ^2}\theta {\sin ^2}\theta
'{\cos ^2}\left( {\phi + \frac{\pi }{4}} \right)\left( {{{\left|
{{C_+}} \right|}^2} - 2\left| {{C_+}} \right|\left| {{C_-}}
\right|\cos {\alpha} + {{\left| {{C_-}} \right|}^2}}
\right).\label{eq:Ec2b}
\end{eqnarray}

Combining Eqs.~(\ref{eq:E0})--(\ref{eq:Ec2b}), we
obtain the energy functional density $\varepsilon = \varepsilon_0
+ \varepsilon_{c_0}+ \varepsilon_{c_2}$.

\vspace{-0.3cm}
\section{Bogoliubov excitations} \label{sec:Bogoliubov} \vspace{-0.3cm}
In this section we derive the equation for the Bogoliubov
excitations discussed in Sec.~\ref{sec:rotons}. The dynamics of a
BEC in the plane-wave or uniform phase is governed by the
time-dependent Gross-Pitaevskii equation
\begin{eqnarray}
i\partial_t\Psi(x,t)=(H_0+H_{\rm{int}})\Psi(x,t), \label{TGP}
\end{eqnarray}
where $H_0$ is the single-particle Hamiltonian in
Eq.~(\ref{eq:Ham2}), $\Psi(x,t)=(\psi_+,\psi_0,\psi_-)^T$, and the
interaction part $H_{\rm{int}}={c_{0}} {\Psi ^{\dag }}\Psi
+{c_{2}}\left( {{\Psi ^{\dag }}\mathbf{F} \Psi }\right) \cdot
\mathbf{F}$.
To calculate the Bogoliubov spectrum of the plane wave
superfluids, we suppose
\begin{eqnarray}
\Psi(x,t)=e^{ikx-i\mu t}(\chi^{(0)}+u_{q}e^{iqx-\omega
t}+v_{q}^*e^{-iqx+i\omega t}), \label{LPsi}
\end{eqnarray}
where $e^{ikx}\chi^{(0)}=
e^{ikx}(\chi^{(0)}_+,\chi^{(0)}_0,\chi^{(0)}_-)^T$ is the ground
state of the system and $u_{q}$ and $v_{q}$ are the wave functions
with three components. Plugging Eq. (\ref{LPsi}) into Eq.
(\ref{TGP}) yields (only keeping linear terms with respect to
$u_q$ and $v_q$)
\begin{eqnarray}
\hbar\omega\left(\begin{array}{c}
{u_{q}}\\
{v_{q}}
\end{array}\right) =\sigma_{z}\mathcal{M}\left(\begin{array}{c}
{u_{q}}\\
{v_{q}}
\end{array}\right),
\end{eqnarray}
where $\sigma_z$ is the $z$-component $2 \times 2$ Pauli matrix,
and
\begin{eqnarray}
\mathcal{M}=\left(\begin{array}{cc}
H_{0}[\hat{p}\rightarrow (k+q)]-\mu+H_{1} & H_{2}\\
H_{2}^{*} & H_{0}^{*}[\hat{p}\rightarrow (k-q)]-\mu+H_{1}^{*}
\end{array}\right),
\end{eqnarray}
with
\begin{eqnarray}
H_{1} & =\left(\begin{array}{cc}
c_{0}(\rho_{0}^{(0)}+|\chi_{+}^{(0)}|^{2})+c_{2}(\rho_{z}^{(0)}+|\chi_{+}^{(0)}|^{2}+|\chi_{0}^{(0)}|^{2}) & c_{0}\chi_{0}^{(0)*}\chi_{+}^{(0)}+c_{2}(\chi_{+}^{(0)}\chi_{0}^{(0)*}+2\chi_{0}^{(0)}\chi_{-}^{(0)*})\\
c_{0}\chi_{+}^{(0)*}\chi_{0}^{(0)}+c_{2}(\chi_{+}^{(0)*}\chi_{0}^{(0)}+2\chi_{0}^{(0)*}\chi_{-}^{(0)}) & c_{0}(\rho_{0}^{(0)}+|\chi_{0}^{(0)}|^{2})+c_{2}(|\chi_{+}^{(0)}|^{2}+|\chi_{-}^{(0)}|^{2})\\
(c_{0}-c_{2})\chi_{+}^{(0)*}\chi_{-}^{(0)} &
c_{0}\chi_{0}^{(0)*}\chi_{-}^{(0)}+c_{2}(2\chi_{+}^{(0)*}\chi_{0}^{(0)}+\chi_{0}^{(0)*}\chi_{-}^{(0)})
\end{array}\right. \nonumber \\
&\left.\begin{array}{c}
(c_{0}-c_{2})\chi_{+}^{(0)}\chi_{-}^{(0)*}\\
c_{0}\chi_{-}^{(0)*}\chi_{0}^{(0)}+c_{2}(2\chi_{+}^{(0)}\chi_{0}^{(0)*}+\chi_{0}^{(0)}\chi_{-}^{(0)*})\\
c_{0}(\rho_{0}^{(0)}+|\chi_{-}^{(0)}|^{2})+c_{2}(|\chi_{-}^{(0)}|^{2}+|\chi_{0}^{(0)}|^{2}-\rho_{z}^{(0)})
\end{array}\right),\\
H_{2} & =\left(\begin{array}{ccc}
(c_{0}+c_{2})\chi_{+}^{(0)2} & (c_{0}+c_{2})\chi_{0}^{(0)}\chi_{+}^{(0)} & (c_{0}-c_{2})\chi_{+}^{(0)}\chi_{-}^{(0)}+c_{2}\chi_{0}^{(0)2}\\
(c_{0}+c_{2})\chi_{+}^{(0)}\chi_{0}^{(0)} & c_{0}\chi_{0}^{(0)2}+2c_{2}\chi_{+}^{(0)}\chi_{-}^{(0)} & (c_{0}+c_{2})\chi_{-}^{(0)}\chi_{0}^{(0)}\\
(c_{0}-c_{2})\chi_{+}^{(0)}\chi_{-}^{(0)}+c_{2}\chi_{0}^{(0)2} &
(c_{0}+c_{2})\chi_{0}^{(0)}\chi_{-}^{(0)} &
(c_{0}+c_{2})\chi_{-}^{(0)2}
\end{array}\right).
\end{eqnarray}
 The Bogoliubov excitation spectrum $\omega$ with
respect to $q$ is numerically obtained by diagonalizing
$\sigma_z\mathcal{M}$ (note that only the physical branches of the
excitations are considered).

\vspace{-0.5cm}

\twocolumngrid


\begin{thebibliography}{99}
\bibitem{Triple_point}
Around $k=0$ one can expand $E_k$ as ${E_0} + \frac{{
f_2}}{2}{k^2} + \frac{{f_4}}{4}{k^4} + \frac{{f_6}}{6}{k^6}$. The
minimum points are at $k_0^2=\frac{{\ - {f_4} + \sqrt {f_4^2
-4{f_2}{f_6}} }}{{2{f_6 }}}$. Therefore, one can obtain the triple
point by solving $f_2(\Omega_{1,2})=f_4(\Omega_{1,2})=0$, which
also determines $E_k -E_0 \sim \frac{f_6}{6} k^6$.
\bibitem{Hellmann-Feynman} One can apply the Hellmann-Feynman theorem to obtain these relations.
\end{thebibliography}
\end{document}